\documentstyle[prl,preprint,aps,epsf]{revtex}
\begin{document}
\draft
\title {Statistical models of nuclear level densities.}
\author{Calvin W.~Johnson$^1$, Jameel-Un Nabi$^1$ and W.~Erich Ormand$^2$ }
\address{
$^1$Department of Physics and Astronomy\\
Louisiana State University,
Baton Rouge, LA 70803-4001\\
$^2$Lawrence Livermore National Laboratory\\
P.O. Box 808, Mail Stop L-414 \\
Livermore, CA 94551
}

\maketitle
\begin{abstract}
We present calculations of nuclear level densities that are based 
upon the detailed microphysics of the interacting shell model 
yet are also computationally tractable.   To do this, we combine 
in a novel fashion several previously disparate ideas from 
statistical spectroscopy, namely partitioning of the model space
into subspaces, analytic calculations of moments up to fourth 
order directly from the two-body interaction, and Zuker's 
binomial distribution.  We get excellent agreement with 
full scale interacting shell model calculations.  We also 
calculate ``{\it ab initio}'' the level densities for 
$^{29}$Si  and $^{57}$Co  
and get  reasonable agreement with experiment. 
\end{abstract}


Reliable nuclear level densities are important for the 
theoretical estimates of nuclear reaction rates in 
nucleosynthesis \cite{Woo78}. The neutron-capture cross sections are approximately 
proportional to the corresponding level densities around the nuclear resonance 
region. The competition between neutron-capture and $\beta$ decay determines 
the fate of 
the {\it s} and {\it r} processes.
%
%
Level densities can be extracted experimentally \cite{Dil73}, 
but reaction network calculations require cross-sections for 
hundreds or thousands of nuclides,
many of them short-lived. 

The most widely used description of the nuclear level density is the 
Bethe formula, based on a gas of free nucleon  \cite{Bet36},
and the  modified ``backshifted Bethe formula''\cite{Hol76}.
Despite its ubiquity the Bethe formula is a phenomenological fit 
 often requiring energy dependent parameters 
to match experimental data.

 The interacting shell model and other 
microscopic models accurately describe spectra and transition for a broad 
range of nuclides.  On the other hand, ``traditional'' shell-model 
codes diagonalize the Hamiltonian in a large-dimensioned basis of 
occupation-state wavefunction but use the Lanczos 
algorithm to  extract only a handful low-lying states. 
The level density requires  complete diagonalization, a 
computationally forbidding requirement.  

An alternative to diagonalization is the Monte Carlo path-integral 
technique \cite{Joh92},
which is well suited to thermal observables \cite{Dea95,Orm97}. 
Although reasonably successful,  path-integral methods are limited 
to interactions that are free of the `sign problem' and 
are still very computationally intensive (i.e., require supercomputer 
time).   
Therefore we feel motivated to consider further alternatives.

In this Letter we combine several previously disparate ideas based 
in nuclear statistical spectroscopy: 
(1) Analytic calculation of moments up to and including fourth-order; 
(2) partitioning the model space into subspaces; and (3) 
using binomial rather than Gaussian distributions.
We find this combination to be  successful,
which we demonstate not only against exact shell-model calculations but 
also against experimental data. 

Nuclear statistical spectroscopy 
argues that many nuclear properties are controlled 
by low-lying moments of the Hamiltonian \cite{Mon75,Won86}. 
The first moment (centroid) is 
$\bar{H} = \mu_1 = 
\left \langle  \hat{H} \right \rangle$; 
for $n>1$ the   $n$th {\it central} moment is
$
\mu_n \equiv \left \langle \left (\hat{H}-\bar{H} \right )^n \right \rangle,
$
We also find it useful to introduce, for $n > 2$, 
 {\it scaled} moments  defined by 
$
m_n \equiv { \mu_n / (\mu_2)^{n/2} }.
$
The width $\mu_2$ provides a natural energy scale.

Analytic formulae exist in the literature for computation of 
centroids through fourth moments directly from two-body matrix elements
 for any number of particles 
without the need for diagonalization in a 
many-body space \cite{Won86,Fre71,Ayi74}. 
Mon and French \cite{Mon75} showed that the level density in a finite space 
tends towards a Gaussian, which is described by the first and second 
moments.  Any accurate description must however include deviations from a 
Gaussian, which require higher moments.  The formulae for third and 
fourth moments \cite{Ayi74} are somewhat time-consuming 
(albeit less so than direct diagonalization and Monte Carlo path 
integration) and so, as far as we can divine from the literature, 
never implemented on a large scale.  Grimes et al \cite{Gri79}
computed higher moments instead by the representative vector method, 
i.e., generating 
random sample wavefunctions in order to estimate the require averages. 
 Today, on a modest workstation a few years 
old, we can compute, for a $0\hbar\omega$ mid-$pf$-shell valence 
space, the third moments in a few hours and 
the fourth moments in a day or two (the time 
for a $0\hbar\omega$ $sd$-shell valence 
space is considerably faster).  While not trivial, we emphasize
such calculations, corresponding to roughly $10^8$ levels,
are still less demanding than Monte Carlo path integration
by a factor of 10  or more in CPU time.

Even with modern computers, however, moments beyond $n=4$ are still 
numerically 
intractable.  Another idea we borrow from statistical spectroscopy is 
partitioning the model space into subspaces, such as single-particle 
configurations $(1d_{5/2})^4 (2s_{1/2})^2$, etc. 
Then, rather than 
computing total moments as defined above, one calculates 
{\it configuration} moments. Let $\alpha$ denote a subspace and let 
$P_\alpha$ be the projection operator for that subspace. 
Then the configuration moment is
$
\mu_n(\alpha) \equiv \left \langle 
P_\alpha \left (\hat{H}-\bar{H} \right )^n \right \rangle.
$
 Configuration moments are applied to configuration or 
partial densities, $\rho_\alpha(E) = {\rm tr \, } 
P_\alpha \delta(E-H)$. 
The total level density  
$\rho(E) = {\rm tr \, } \delta(E-H)$ is simply the sum of 
the partial densities.  

Partitioning has several advantages. First, it comes at no  
cost: the formulae in the literature for total moments are already expressed 
in terms of sums of configuration moments, which are thus needed and readily 
available.  Second, higher total moments ($n > 4$) are dominated 
by lower configuration moments. Finally, 
partial densities themselves are useful for calculations
of preequilibrium emission in compound nuclei \cite{Plu88,Dea99}.
 
The final step is a ``base model'' for the level density.  Essentially, 
this base model makes assumptions for the values of the higher moments 
given the lower moments.  The most common base model in statistical 
spectroscopy is the Gaussian distribution
 and its various modifications. (The random matrix 
model of Pluhar and Weidenm\"uller \cite{Plu88} 
is built upon semicircle distributions, 
which has both advantages and disadvantages that we do not have the space 
to discuss here.)  A Gaussian is reasonable good starting point as  
it is already close to the actual level density. 
A common generalization is to expand in a Gram-Charlier series 
using Hermite polynomials \cite{Won86}; this unfortunately 
can lead to negative level densities.  Another generalization is to 
extend the Gaussian to a function of the form 
$\exp(- \alpha E^2 - \beta E^3- \gamma E^4 \ldots ) $  
\cite{Cha72,Gri83} but the relation 
between  parameters 
$\alpha , \beta , \gamma $ and the moments is not amenable to 
a simple analytic formula.

Instead we chose to  follow Zuker, who recently gave a combinatorial 
argument that one should use 
binomial rather than  Gaussian distributions 
to approximate level densities \cite{Zuk99}.  
Zuker showed how third moments, that is,  asymmetric distributions, 
are easily handled by binomials.  Although Zuker 
did not comment on fourth moments, we  find that
the fourth moment  can also be controlled in binomials.

Consider the binomial expansion 
\begin{equation}
(1+\lambda)^N = \sum_{k = 0}^N \lambda^k.
{N \choose k}
\end{equation}
Now interpret this binomial expansion as the density of states. 
At the excitation energy 
$E_x = \epsilon k$, $\epsilon$ being an 
overall energy scale, the number of states is  
$\lambda^k {N \choose k}$ .
(Shortly we will see that deviations of $\lambda$ from 1 represent
an asymmetric distribution.) 
Because we can write ${N \choose k}$ with gamma functions, one 
can easily approximate it by a continuous distribution,
\begin{equation}
\rho(E_x) = \lambda^{E_x / \epsilon} 
{ \Gamma(E_{max}/\epsilon+1) \over \Gamma(E_x / \epsilon +1) 
\Gamma( (E_{max} - E_x) /\epsilon +1 )}
\end{equation}
where $E_{max} = N\epsilon$ represents exhaustion of the finite number
of states. Although we began with $N$ as an integer, 
it no longer has to be.

The total number of states, which is the `zeroth' moment, is 
$
d= (1+\lambda)^N,
$
while the centroid and 
width are given respectively by 
\begin{eqnarray}
\mu_1 = {N \epsilon \lambda \over 1+\lambda}
\label{moment1} \\
\mu_2 = {N \epsilon^{2} \lambda \over (1+ \lambda)^{2}} 
\end{eqnarray}
and the {\it scaled} third and fourth moments are 
\begin{eqnarray}
m_{3}={1- \lambda \over \sqrt{N \lambda}},
\label{moment3}\\
m_4 =
3- {4-\lambda \over N}+ {1 \over N \lambda}.
\label{moment4} 
\end{eqnarray}
For Gaussians ($N \rightarrow \infty $) $m_4=3$. For 
most binomials and for shell model diagonalization, the scaled fourth moment 
is less than 3, a typical value being around 2.8. In random 
matrix theory, the semicircle distribution typical of Gaussian Orthogonal 
Ensembles has $m_4 =2 $.

Using Stirling's approximation, and a few others, 
Zuker arrives at 
\begin{equation}
\rho(E_{x}) \approx \sqrt{\frac{8}{N \pi}} 
\exp \left (
-(N-1)\left (\frac{E_{x}}{E_{max}}\ln{\frac{E_{x}}{E_{max}}}+ 
\frac{E_{max}-E_{x}}{E_{max}}\ln{\frac{E_{max}-E_{x}}{E_{max}}}\right 
)+N\frac{E_{x}}{
E_{max}}\ln{\lambda} \right ).
\label{density}
\end{equation}
We carefully note that the above moments (\ref{moment1}-\ref{moment4})
are exact for 
discrete distributions but are only approximate for the 
continuous distribution (\ref{density}). 
For large $N$, however, they are very good approximations.

The key parameters of the binomial are the order $N$ and the asymmetry 
parameter $\lambda$.  In the 
limit $\lambda =1$ and  $N\rightarrow \infty$ one regains the Gaussian. 
 If $\lambda=1$ then the binomial is  symmetric: 
 $m_3=0$; if $\lambda \neq 1$ then the binomial is  asymmetric.  
Zuker suggested that the order of the binomial, $N$, be fixed 
by the dimension of the model space. In that case $N$ and $\lambda$ 
are fixed by solving $d=(1+\lambda)^N$ and Eqn.~(\ref{moment3}) 
simultaneously.  This we consider as the standard binomial, which 
can be asymmetric (nonzero third moment).  

We observe,  
however, that one could instead fix the order $N$ by the fourth moment, 
and solve (\ref{moment3}) and (\ref{moment4}) simultaneously instead,
afterwards multiplying the entire binomial distribution by a constant 
so as to get the correct total number of levels. 
This we refer to as the {\it fourth-moment scaled} (FMS) binomial. 
After $N$ and $\lambda$ are determined, the centroid and width simply 
fix the absolute position and scale of the distribution. 

Elsewhere \cite{Nab01} we compare in detail the relative importance of third 
and fourth moments and configuration versus total moments.  
We find that the level density is best described by 
a sum of partitioned binomials, and that the difference between 
this best description and others is, at low energy, a factor of two or 
more.   In the rest of this Letter we compare this statistical approach 
with exact shell model calculations, both from direct diagonalization and 
from Monte Carlo path integration, and with experimental data.  Not only do we 
reproduce the secular behavior of the level density, we find that in some 
cases the partitioned model space can somewhat 
describe detailed structures at low excitation energy if one uses 
FMS binomials.

To test the approach outlined above
 we considered a number of $sd$- and $pf$-shell 
nuclides, and show only a representative sample here; more can be 
found in \cite{Nab01}.  What we plot is in fact the {\it state} density, 
which includes  $2J+1$ degeneracies.  Strictly speaking, the {\it level 
density} ignores $2J+1$ degeneracies, but the literature is often cavalier 
with this distinction.  

First we consider  full $0\hbar\omega$ $sd$-shell calculations,
using the Wildenthal USD interaction \cite{Wil84} 
and focus on  several nuclides  that could 
be completely diagonalized using the OXBASH shell-model code \cite{oxbash}. 
The same single-particle energies and two-body interaction 
matrix elements were used by OXBASH for the exact calculation 
and the routines to compute the 
configuration moments for our statistical approximations. 
Fig.~1 shows three typical cases: $^{32}$S, $^{24}$Mg and $^{22}$Na. 
The histograms are the exact state densities.  We consider 
these three cases because they display very disparate 
collective behaviors: vibrational, 
rotational, and noncollective, respectively.   
While the standard binomial does reasonably well, 
the sum of partitioned FMS binomials describe all three somewhat 
better, despite different collective behaviors. We attribute the 
minimal differences for $^{22}$Na to its noncollective, fully 
statistical behavior even at the lowest energies. The curious 
tracking by the summed FMS binomials of low-lying structure in
the state density is very intriguing, although such tracking,
as well as the structure in the exact histograms, is 
less pronounced in $pf$-shell nuclides. (NB: that all three examples 
are $N=Z$ is irrelevant; the standard binomial and partitioned 
binomials work as well for $N \neq Z$ although Gaussians fair 
even worse, because of more pronounce asymmetries (larger $m_3$).)

We have also performed $0\hbar\omega$ calculations in the 
$pf$-shell. 
Figure 2 shows  results for $^{54}$Fe and $^{48}$Cr.  
Because of the prohibitively large dimensions of this model space,
these nuclides cannot be   
diagonalized to yield all eigenvalues. 
Instead we turned to Monte Carlo path integration.
To avoid the well-known sign problem \cite{Joh92} we fitted a schematic 
multipole-multipole interaction  to the $T=1$ matrix elements 
of the FPD6 interaction of Richter et. al \cite{richter}.  
Once again the sum of binomials is clearly superior to either a Gaussian 
or a single standard binomial. (It turns out that one needs 
a sum of binomials rather than a sum of Gaussians because the 
third configuration moments contribute nontrivially. Summing 
symmetric binomials also yields poor results.) We note 
that here the difference between the sum of 
``standard  binomials'' and sum of 
FMS binomials is negligible; furthermore  there is much less low-lying 
structure in the state density than for $sd$-shell nuclides.

Finally, in Figure 3 we compare directly to experimental data, 
for $^{29}$Si \cite{si29} and $^{57}$Co \cite{co57}.  
Here it was necessary to include 
odd-parity states. For $^{29}$Si we used the Wildenthal interaction 
\cite{Wil84} in the $sd$-shell; to account for excitations 
out of the $sd$-shell we included particle-hole excitations 
into the $pf$ shell using the WBMB interaction of 
Ref.~\cite{wbmb}.
 To avoid the ``$N$-$\hbar\omega$ catastrophe'', we follow the 
suggestion of Ref.~\cite{wbmb} and decouple the $n\hbar\omega$ excitations.
Although clearly one should worry about contamination by spurious 
center-of-mass motion, we temporarily put this concern aside 
(as do all Monte Carlo calculations to date). 
Similarly, for $^{57}$Co we used the 
modified KB3 interaction \cite{KB3} and allowed 1-particle, 1-hole 
excitations into a noninteracting $g_{9/2}$ orbit whose single-particle
energy is set by the start of abnormal parity states.  
Our calculations fall somewhat short at high energy, 
but this is clearly due to insufficient particle-hole states 
in our calculation.  The overall agreement with experiment is 
remarkable, especially considering that the interactions were 
tuned to very low-lying states and not to more global properties such 
as state densities.  While our treatment of particle-hole states 
here is admittedly ad hoc, keep in mind that  $n\hbar\omega$ excitations 
is always something of an art in shell model calculations.
Indeed, we believe that our microscopic calculations are limited 
more by the uncertainty in the appropriate interaction than 
our statistical approximations, and that this should be the focus 
of future investigations.

In summary: we have outlined a theoretical approach to level densities 
that is both microscopic in origin and also computationally tractable. 
Application to higher shells is hampered as much by our ignorance of the 
effective two-body interaction as anything else.  In the near term we will 
also work to extend this approach to calculation of spin-cutoff factors 
and estimate of contamination by spurious states.

This work was performed under the auspices of the 
Louisiana Board of Regents, 
contract number LEQSF(1999-02)-RD-A-06; and under the auspices of the U.S. Department of Energy 
through the University of California, Lawrence Livermore National Laboratory, 
under contract No. W-7405-Eng-48.

\begin{figure}[h!]
\epsfxsize=13.0cm
\epsffile{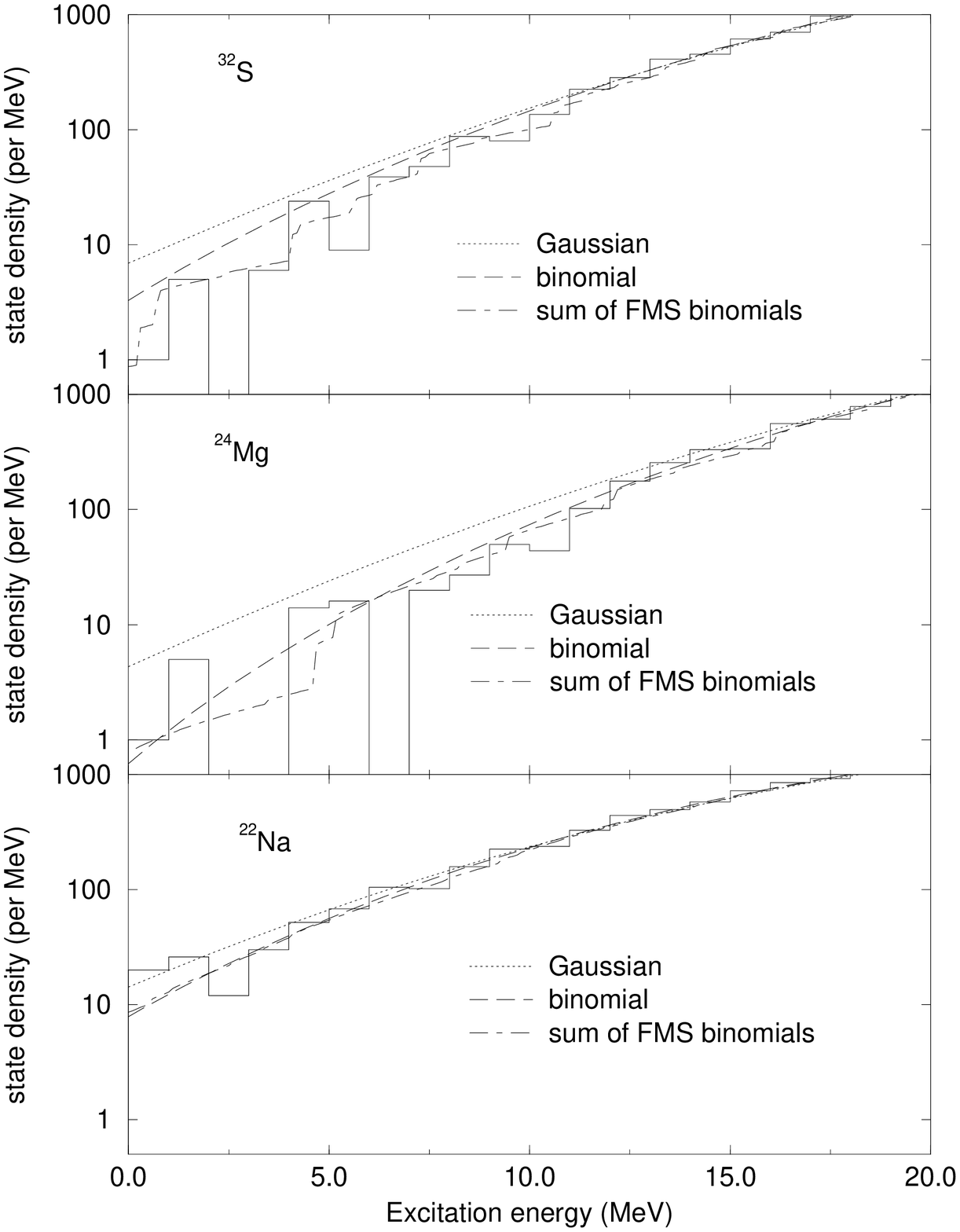}
\caption{ Comparison of exact shell model state densities  (histograms)
from direct diagonalization in 
a full $0\hbar\omega$ basis against statistical approximations for 
selected $sd$-shell nuclides. FMS = ``fourth moment scaled'' (see 
text).
}
\end{figure}

\begin{figure}[h!]
\epsfxsize=13.0cm
\epsffile{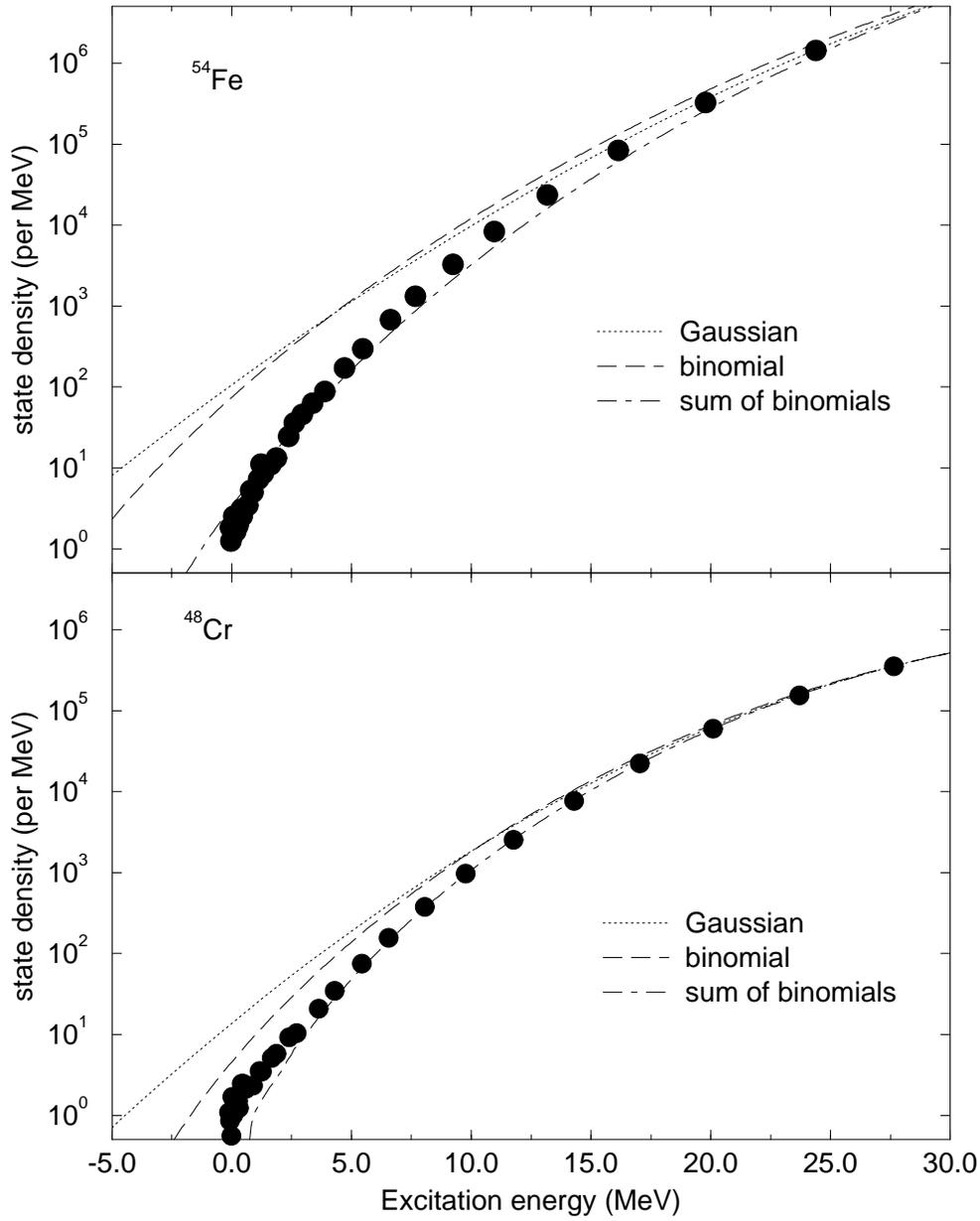}
\caption{ Comparison of exact shell model state densities  (circles)
from Monte Carlo path integration in 
a full $0\hbar\omega$ basis against statistical approximations for 
selected $pf$-shell nuclides.
}
\end{figure}

\begin{figure}[h!]
\epsfxsize=13.0cm
\epsffile{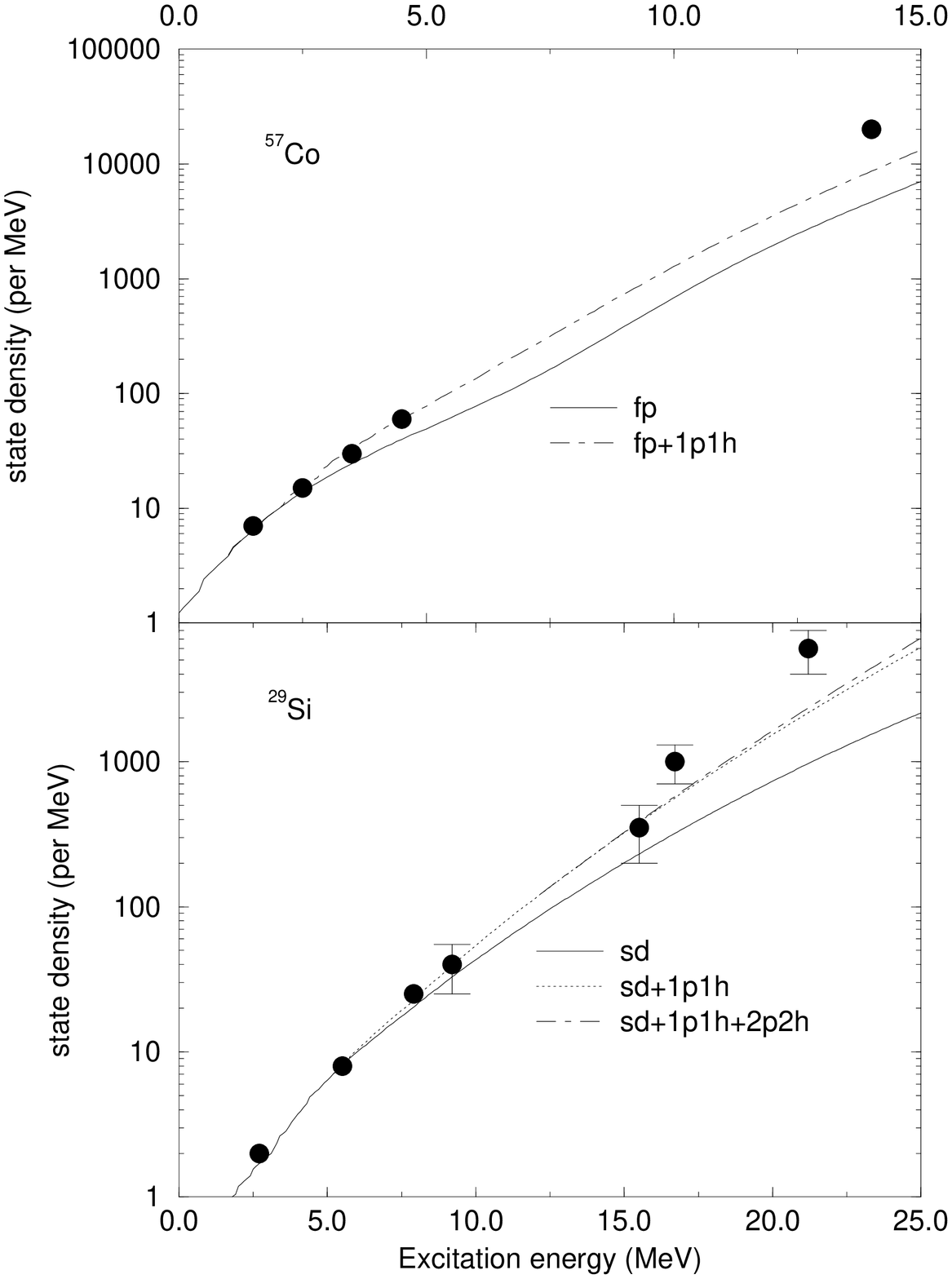}
\caption{ Comparison of experimental data of state densities 
for $^{57}$Co and $^{29}$Si against a sum of binomials. 
Error bars in the lower figure represent the spread in experimental 
data. 
}
\end{figure}


\begin{thebibliography}{99}
\bibitem{Woo78}S.E. Woosley, W.A. Fowler, J.A. Holmes and B.A. Zimmerman, At. 
Nucl. Data Tables {\bf 22}, 371 (1978);
F.-K. Thielemann, M. Arnould and J.W. Truran, Advances in 
Nuclear Astrophysics, eds. E. Vangioni-Flam et. al. p.525.
%
%
\bibitem{Dil73}W. Dil, W. Schantl, H. Vonach and M. Uhl, Nucl. Phys. A {\bf 
217}, 269 (1973);
J.R. Huizenga, H.K. Vonach, A.A. Katsanos, A.J. Gorski and C.J. 
Stephan, Phys. Rev. {\bf 182},1149 (1969);
C.C. Lu, L.C. Vaz and J.R. Huizenga, Nucl. Phys. A {\bf 190}, 
229 (1972);
A. Schiller, L. Bergholt, M. Guttormsen, E. Melby, J. Reststad 
and S. Seim, Nucl. Instum. Methods A {\bf 447}, 498 (2000).

\bibitem{Bet36}H.A. Bethe, Phys. Rev. {\bf 50}, 332 (1936).
%
%
\bibitem{Hol76}J.A. Holmes, S.E. Woosley, W.A. Fowler and B.A. Zimmerman, 
Atom. Data, Nucl. Data Tables {\bf 18}, 305 (1976);
J.J. Cowan, F.-K. Thielemann and J.W. Truran, Phys. Rep. {\bf 
208}, 267 (1991).


\bibitem{Joh92}C.W. Johnson, S.E. Koonin, G.H. Lang and W.E. Ormand, Phys. 
Rev. Lett.  {\bf 69}, 3157 (1992).
\bibitem{Dea95}D.J. Dean, S.E. Koonin, K. Langanke, P.B. Radha and Y. 
Alhassid, Phys. Rev. Lett. {\bf 74}, 2909 (1995).

\bibitem{Orm97} W.~E.~Ormand,  Phys.~Rev.~ C 
{\bf 56}, R1678 (1997);  H.~Nakada and Y.~Alhassid,
 Phys.~Rev.~Lett. {\bf 79} (1997) 2939;
H. Nakada and Y. Alhassid, Phys. Lett. B {\bf 436}, 231 (1998);
J.A. White, S.E. Koonin and D.J. Dean, Phys. Rev. C {\bf 61}, 
034303 (2000).


\bibitem{Mon75} K. K. Mon and J. B. French, Ann. Phys. {\bf 95}, 90 (1975).

\bibitem{Won86}  S.~S.~M.~Wong, {\it Nuclear Statistical Spectroscopy},
Oxford Press (New York, 1986).

\bibitem{Fre71}J.B. French and K.F. Ratcliff, Phys. Rev. C {\bf 3}, 94 (1971)
\bibitem{Ayi74}S. Ayik and J.N. Ginocchio, Nucl. Phys. A {\bf 221}, 285 (1974)



\bibitem{Gri79}S. M. Grimes, S. D. Bloom, R. F. Hausman, Jr. and B. J. 
Dalton, Phys. Rev. C {\bf 19}, 2378 (1979)
%

%
\bibitem{Cha72}F. S. Chang and A. Zuker, Nucl. Phys. A {\bf 198}, 417 (1972)


\bibitem{Gri83}S. M. Grimes, S. D. Bloom, H. K. Vonach and R. F. Hausman, 
Jr., Phys. Rev. C {\bf 27}, 2893 (1983)

%
\bibitem{Plu88} Z.~Pluhar and H.~A.~Weidenm\"uller, {\it Phys.~Rev.~C }
{\bf 38} (1988)  1046.

\bibitem{Dea99} D. J. Dean and S. E. Koonin, Phys. Rev. C {\bf 60}, 
054306 (1999).

\bibitem{Zuk99}A. P. Zuker, LANL archive 
nucl-th/9910002.

\bibitem{Nab01} J.~Nabi, C.~W.~Johnson, and W.~E.~Ormand, 
to be published. 


\bibitem{Wil84}
B.H. Wildenthal, Prog. Part. Nucl. Phys. {\bf 11}, 5 (1984).

\bibitem{oxbash} B.~A.~Brown, A.~Etchegoyen, and W.~D.~M.~Rae, OXBASH,
the Oxford University-Buenos Aires-MSU shell model code, Michigan State
University Cyclotron Laboratory Report No. 524 (1985).

\bibitem{richter} W.~A.~Richter, M.~G. van der Merwe, R.~E.~Julies, and
B.~A.~Brown, Nucl. Phys. {\bf A523}, 325 (1990).

\bibitem{si29}
F.B. Bateman, S. M. Grimes, N. Boukharouba, V. Mishra, C.E. Brient, R.S. 
Pedroni, T.N. Massey and R.C. Haight, Phys. Rev. C {\bf 55}, 133 (1997).

\bibitem{co57} 
V. Mishra, N. Boukharouba, C.E. Brient, S. M. Grimes, and R.S. Perdoni, 
Phys. Rev. C {\bf 49}, 750 (1994).

\bibitem{wbmb} E.~K.~Wharburton, J.~A. Becker, and B.~A.~Brown, Phys. Rev.
{\bf C41}, 1147 (1990).

\bibitem{KB3} T.~Kuo and G.~E.~Brown, Nucl.~Phys.~{\bf A114}, 241 (1968); 
A.~Poves and A.~P.~Zuker, Phys.~Rep. {\bf 70}, 235 (1981).
\end{thebibliography}
\end{document}